\documentclass[final,11pt,5p,times,twocolumn,sort&compress]{elsarticle}

\usepackage{epstopdf}
\usepackage{graphicx}
\usepackage{multicol}
\usepackage{enumerate}
\usepackage{tabularx}
\usepackage{rotating}
\usepackage{anysize}
\usepackage{subfigure}
\usepackage{float}
\usepackage{hyperref}

\def\TeV{\ifmmode {\mathrm{\ Te\kern -0.1em V}}\else
                   \textrm{Te\kern -0.1em V}\fi}%
\def\GeV{\ifmmode {\mathrm{\ Ge\kern -0.1em V}}\else
                   \textrm{Ge\kern -0.1em V}\fi}%
\def\MeV{\ifmmode {\mathrm{\ Me\kern -0.1em V}}\else
                   \textrm{Me\kern -0.1em V}\fi}%
\def\keV{\ifmmode {\mathrm{\ ke\kern -0.1em V}}\else
                   \textrm{ke\kern -0.1em V}\fi}%
\def\eV{\ifmmode  {\mathrm{\ e\kern -0.1em V}}\else
                   \textrm{e\kern -0.1em V}\fi}%
\let\tev=\TeV
\let\gev=\GeV

\begin{document}

\begin{frontmatter}
\vskip -2cm
\title{A new jet reconstruction algorithm for lepton colliders}
\author{M. Boronat}
\author{I. Garc\'ia} 
\author{M. Vos} \address{IFIC (CSIC/UVEG), Valencia, Spain}

\date{\today}

\begin{abstract}
We propose a new sequential jet reconstruction algorithm for future lepton
colliders at the energy frontier. The Valencia algorithm combines 
the natural distance criterion for lepton colliders with the greater
robustness against backgrounds of algorithms adapted to hadron colliders.
Results on a detailed Monte Carlo simulation of $t\bar{t}$ and $ZZ$ production
at future linear $e^+e^-$ colliders (ILC and CLIC) with a realistic level 
of background overlaid, show that it achieves better 
performance in the presence of background.
\end{abstract}


\end{frontmatter}

\section{Introduction}
\label{sec:intro}


Experiments at lepton and hadron colliders use jet algorithms to 
cluster the collimated sprays of particles that form in processes 
with asymptotically free quarks and gluons in the final state. The first modern 
sequential recombination algorithms were developed for $e^+ e^-$ colliders 
operated at the $Z$-pole (a detailed historical account is 
found in Reference~\cite{Moretti:1998qx}).
At the heart of the jet algorithm - and crucial to the definition of jets
themselves - is a criterion to define the distance between two particles.  
In popular algorithms used at $e^+e^-$ colliders the distance combines 
information on the angle between the particles and the energy of 
(the softest of the two) particles.
Sequential recombination algorithm were adapted to the environment at 
hadron colliders in the early 1990s. At the Large Hadron Collider 
the large majority of analyses is based on inclusive jet reconstruction 
with the anti-$k_t$ algorithm~\cite{Cacciari:2008gp}.

An intense R\&D programme exists to develop the technology required for 
an $e^+e^-$ collider with a center-of-mass energy well beyond that of previous
lepton colliders. A linear $e^+e^-$ collider 
can attain center-of-mass energies from several 100~\gev{} to 
several~\tev{}~\cite{Baer:2013cma,Linssen:2012hp}. 
The possibility of a large circular $e^+e^-$ collider 
that can reach a center-of-mass energy of approximately 
350~\gev{}~\cite{Gomez-Ceballos:2013zzn} is also explored,
as well as a muon collider~\cite{Alexahin:2013ojp}. 
Such machines present an environment that differs in
several important respects from that encountered at the Z-pole. 
In this Letter we explore which jet reconstruction algorithms 
are most suitable for the $e^+e^-$ colliders with a center-of-mass 
energy from 100~\gev{} to several~\tev.

We start our discussion with a brief recapitulation of the properties of 
the most popular clustering algorithms in Section~\ref{sec:clustering}. 
We present a proposal for a new jet 
algorithm in Section~\ref{sec:vlc}. In Section~\ref{sec:comparison} 
the key features of this algorithm are compared to popular algorithms.
In Section~\ref{sec:simulation} the Monte Carlo simulation
setup that we used to benchmark the performance of the algorithms
is introduced.
Finally, in Sections~\ref{sec:resultsilc500} and~\ref{sec:resultsclic500}
we present the results for top quark pair and 
di-boson ($ZZ$) production at the ILC and CLIC, in a realistic
environment including the relevant background. In 
Section~\ref{sec:conclusions} we summarize the most
important findings of this work.

\section{Overview of jet reconstruction algorithms based on sequential recombination}
\label{sec:clustering}

The first modern clustering algorithm with a simple sequential recombination 
scheme algorithm is the JADE algorithm developed in the middle of the 
1980’s~\cite{Bartel:1986ua,Bethke:1988zc}. The distance $y_{ij}$ assigned
to any pair of particles $i$ and $j$ is given by:
\begin{equation} 
y_{ij} = \frac{E_i^2 E_j^2}{Q^2} (1 - \cos{\theta_{ij}})
\end{equation}
where $E_i$ and $E_j$ denote the energy of the two particles, $Q$ is the 
total energy of the event, and $\theta_{ij}$ is the angle between the two
particles. 
At each step the algorithm merges the pair of particles with the smallest
distance $y_{ij}$. This process continues until the smallest distance
exceeds a value $y_{cut}$ ({\em inclusive} clustering) or a previously
defined number of jets is obtained ({\em exclusive} clustering).

In the Durham or $e^+e^-$ $k_t$ algorithm~\cite{Catani:1991hj} used
extensively at LEP and SLC the distance between particles $i$ and $j$ 
is modified to depend on the minimum of the energies $E_i$ and $E_j$, rather 
than the product $E_iE_j$:
\begin{equation} 
d_{ij} = 2 min(E_i^2,E_j^2) (1 - \cos{\theta_{ij}})
\end{equation}
For sufficiently small angles the distance reduces to the transverse momentum
squared of the softer particle relative to the harder one. The distance
measure is thus proportional to the squared inverse of the splitting 
probability for one parton $k$ into partons $i$
and $j$ in the soft and collinear limit.

Jet reconstruction at hadron colliders presents 
a number of additional difficulties. The incoming beams radiate gluons
that can form jets. Only a fraction of the energy of the composite 
projectiles is transferred in the hard parton-parton process
and a hadron remnant continues to travel down the beam pipe. 
An important consequence is that the system formed by the reaction
products is typically not at rest in the laboratory frame\footnote{ 
For di-jet production at the LHC $\beta_z = v_z/c$ of the di-jet system
is very close to 1 and even a massive system such as a top quark pair
acquires a typical $\beta_z =$ 0.5. In contrast, for processes such 
as $e^+e^- \rightarrow ZH (\gamma)$ (Higgsstrahlung) at $\sqrt{s} =$ 250~\gev{} 
and $e^+e^- \rightarrow t \bar{t} (\gamma)$ at 500~\gev{} $\beta_z$ is smaller
than 0.1 in 95\% and 90\% of the events, respectively. 
The exception to the rule is the $2 \rightarrow 2 $ process 
$e^+e^- \rightarrow f \bar{f} (\gamma)$, with $f$ any fermion lighter than 
the $Z$-boson, where ISR (return-to-the-Z) plays an important role.}.
Clustering algorithms were adapted to meet these challenges in the 1990s.

The first important modification of the algorithms 
is the addition of so-called {\em beam jets}, introduced in
Reference~\cite{Catani:1992zp}. 
Any particle with a beam distance $d_{iB} = p_{Ti}^{2n}$ smaller than
any $d_{ij}$ is not merged with any other particle, but is associated 
to the beam jet. These are not considered part of the visible final state.
Thus, the soft, collinear radiation emitted by the incoming hadrons 
and the hadron remnant travelling in the very forward
and backward direction are discarded.

To cope with the boost along the beam direction, analyses at hadron
colliders replace the particle energy $E_i$ with its transverse 
momentum $p_{Ti}$ and the angular distance between the 
particles $(1 - \cos{\theta_{ij}})$
with $\Delta R_{ij} =  \sqrt{(\Delta \phi)^2 + (\Delta y)^2}$, where
$y$ denotes the rapidity. In the longitudinally invariant $k_t$ 
algorithm~\cite{Catani:1993hr,Ellis:1993tq} the distance criterion is based
on the same observables  
``to improve the factorization properties [of the algorithm] and [achieve] 
closer correspondence to experimental practice [...]''~\cite{Catani:1993hr}.
We rewrite the generic inter-particle distance as follows:
\begin{equation}
d_{ij} = min(p_{Ti}^{2n}, p_{Tj}^{2n}) \frac{\Delta R_{ij}^{2}}{R^{2}}
\end{equation}
where $R$ is the radius parameter.
Setting $n$ in the exponent to 1 yields the longitudinally invariant
$k_t$ algorithm. Alternative choices of the exponent yield the 
Cambridge-Aachen algorithm ($n=$0), or the anti-$k_t$ algorithm ($n=$-1), 
the default jet reconstruction algorithm at the LHC.

Finally, one can add beam beam jets to the $k_t$ algorithm for $e^+e^-$ 
experiments. This yields an algorithm we refer to as the generic 
$e^+e^-$ $k_t$ algorithm, with inter-particle distance:
\begin{equation}
 d_{ij} = min(E_i^2,E_j^2) (1 - \cos{\theta_{ij}})/(1 - \cos{R})
\label{eq:eekt}
\end{equation}
and beam distance given by $d_{iB} = E_i^2$.

\section{The Valencia jet algorithm}
\label{sec:vlc}

Background levels at hadron colliders form an important consideration
in the design of jet algorithms. The {\em pile-up} of several tens
of minimum bias events on each bunch crossing at the LHC is
a serious challenge that has led to a large body of work on
mitigation and correction methods. In comparison, previous lepton colliders, 
such as LEP or SLD, presented an environment with essentially negligible 
background. Future lepton colliders are in between these two extremes.
While very far from the background levels 
of the LHC, detailed studies of the $\gamma \gamma \rightarrow$ {\em hadrons}
background at the ILC or CLIC have shown a non-negligible impact on the jet 
reconstruction performance~\cite{Linssen:2012hp,Marshall:2012ry}.
Among several proposal to mitigate its effect, the use of the
longitudinally invariant $k_t$ algorithm, intended for hadron colliders, 
has led to the greatest improvement of the robustness.

We propose a new clustering jet reconstruction algorithm for future $e^+e^-$ 
colliders that maintains a Durham-like distance criterion based on 
[energy, polar angle] (as opposed to [transverse momentum, rapidity] in
the hadron collider algorithm) and can compete with the robustness 
against background of the longitudinally invariant $k_t$ 
algorithm. The algorithm has the following inter-particle distance:
\begin{equation} 
 d_{ij} = min(E_i^{2 \beta},E_j^{2 \beta}) (1 - \cos{\theta_{ij}})/R^2
\end{equation}
For $\beta = $1 the distance is given by the transverse momentum squared of 
the softer of the two particles relative to the harder one, as in the 
Durham algorithm. Note that we have redefined the meaning of the 
radius parameter $R$ with respect to the generalized $e^+e^-$ algorithm
with beam jets. The $R^2$ in the numerator yields greater freedom than 
the $1 - \cos{R}$, that is limited to the interval $[0,2]$. 

The beam distance of the Valencia algorithm is:
\begin{equation}
 d_{iB} = p^{2 \beta}_T
\end{equation}

For $\beta=$ 1 this combination of inter-particle and beam distance metrics 
is similar to that of the $k_\perp$ algorithm proposed in 
Ref.~\cite{Catani:1992zp}, with the difference that 
$d_{iB} = p_{ti}^2 = E_i^2 \sin^2{\theta_{iB}}$, whereas in Ref.~\cite{Catani:1992zp} it was given by $2 E_i^2 (1 - \cos{\theta_{iB}})$. 


The Valencia algorithm is available as a plug-in for the
FastJet~\cite{Cacciari:2005hq,Cacciari:2011ma} package. 
The code can be obtained from the ``contrib'' area~\cite{fjcontrib}.

\section{Comparison of the distance criteria of sequential recombination algorithms}
\label{sec:comparison}

\begin{figure*}[htbp!]
  {\centering 
\includegraphics[width=0.32\textwidth]{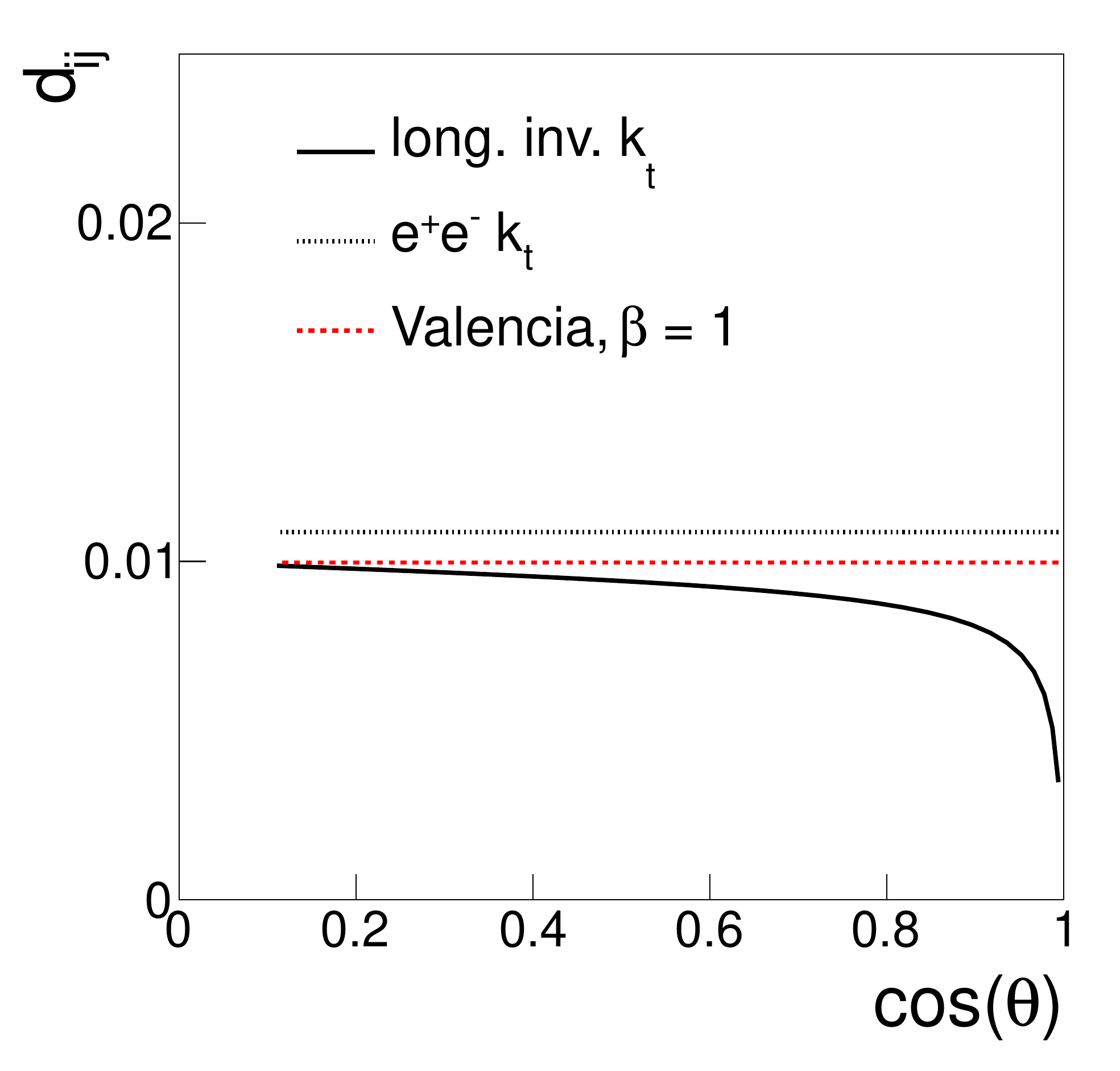}
\includegraphics[width=0.32\textwidth]{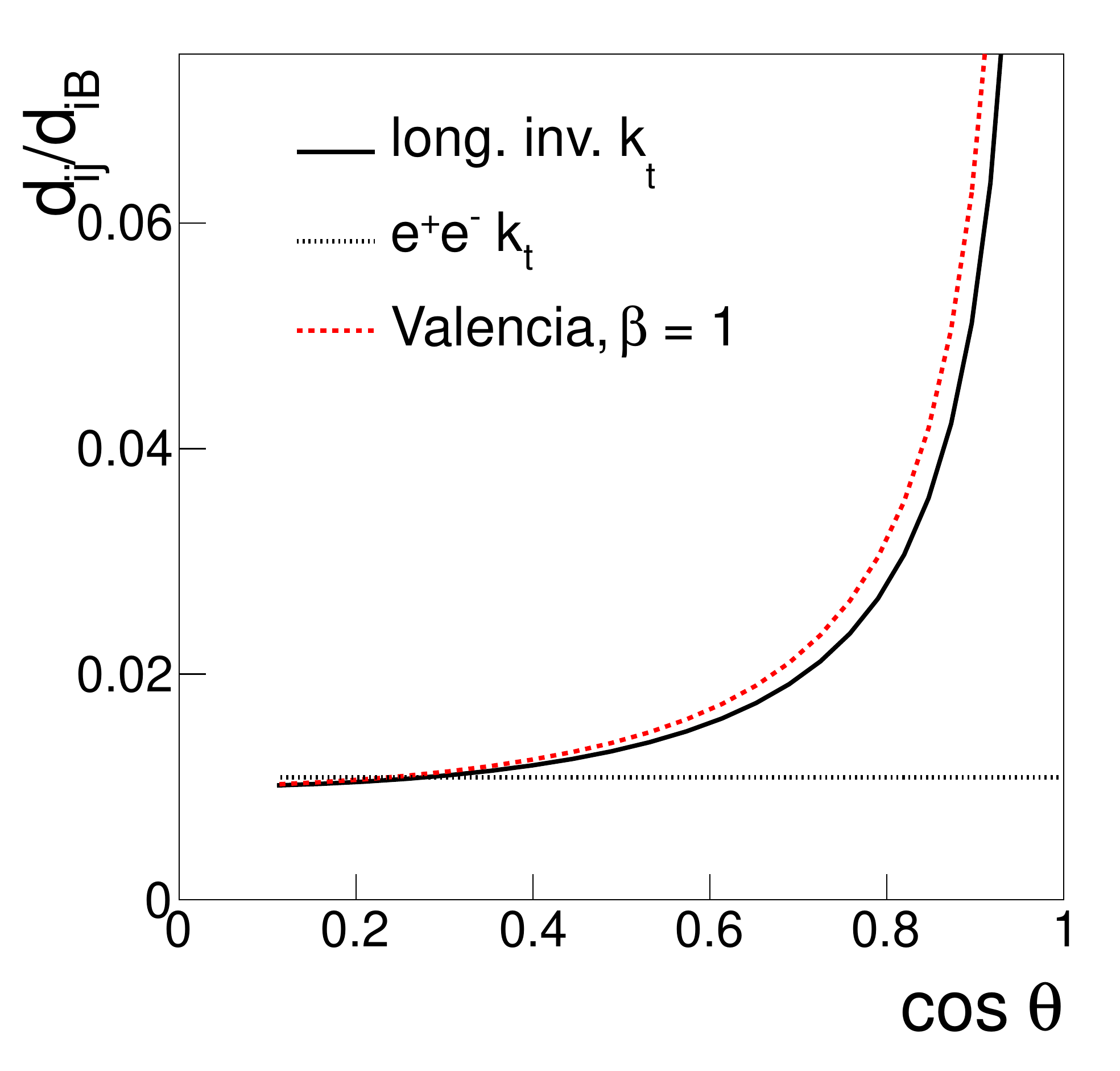}
\includegraphics[width=0.32\textwidth]{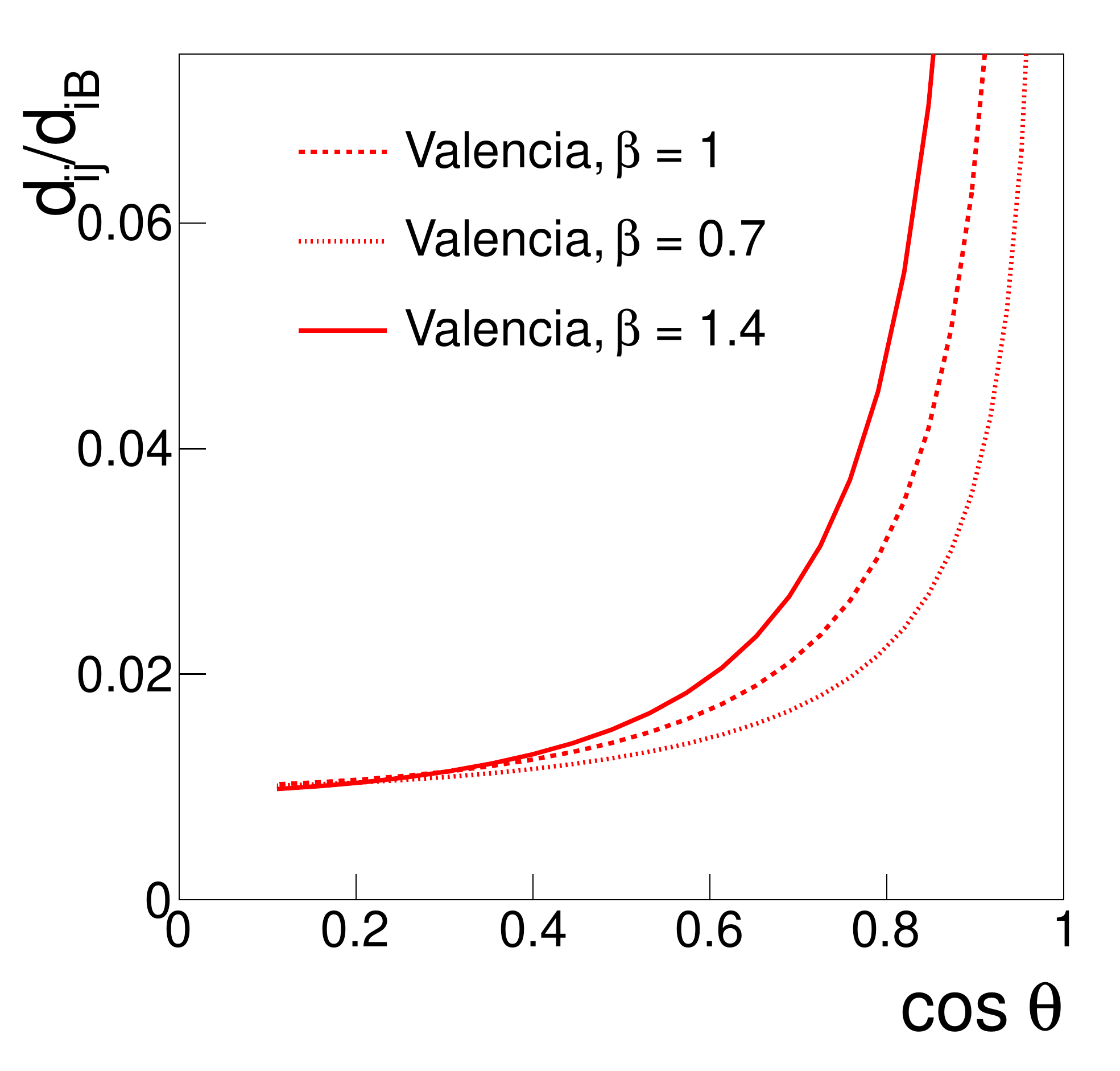}
\caption{The dependence of the inter-particle distance $d_{ij}$ of two test particles emitted at fixed angular distance and the ratio of $d_{ij}$ to the beam
distance $d_{iB}$ with the polar angle $\theta$. Results are presented for several clustering jet reconstruction algorithms discussed in the text.}
	\label{fig:distance_vlc}
	}
	\end{figure*}

The choice of distance criterion defines the essence of the jet algorithm
and has profound implications on its performance in a given environment.
The differences between the various algorithms are most easily visualized 
as follows. We calculate
the distance between two test particles with an energy of 1~\gev{} emitted
at a fixed relative angle of 100 mrad. The leftmost plot in 
Figure~\ref{fig:distance_vlc} shows how the distance between the two 
particles evolves as the system is scanned from the central detector 
($\cos{\theta} = 0$) to the forward region ($\cos{\theta} = 1$). 

The distance $d_{ij}$ of the generic $e^+e^-$ $k_t$ algorithm is independent 
of polar angle, as shown in Figure~\ref{fig:distance_vlc}. The same  
holds for the Valencia algorithm proposed here, but generally not for 
algorithms used at hadron colliders. Two effects come into play. 
For two particles separated by 
a given polar angle, the pseudo-rapidity difference $\Delta \eta$ grows larger 
in the forward region. At the same time the
distance between two particles with energy $E$ decreases as $p_T$ is
reduced. The net effect for the $k_t$ algorithm is a sharp decrease
of the distance in the forward region.

The relation between the inter-particle distance $d_{ij}$ and the beam 
distance $d_{iB}$ governs the relative {\em attraction} of beam jets and 
final-state jets and is therefore a crucial property for the performance 
in environments with significant background. 
The ratio $\frac{d_{ij}}{d_{iB}}$ is shown as a function of polar angle in 
the central plot in Figure~\ref{fig:distance_vlc}. 
As might be expected from the functional form in Equation~\ref{eq:eekt}, 
the ratio is flat for $e^+e^-$ algorithms (Durham). 
For the longitudinally invariant
$k_t$ algorithm, on the other hand, the ratio rises steeply in the 
forward region. For the Valencia algorithm with $\beta=1$ we obtain very
similar behaviour to longitudinally invariant $k_t$. 


The steep rise in $\frac{d_{ij}}{d_iB}$ at $\cos{\theta} \sim 1$ penalizes 
relatively isolated particles in the forward and backward directions, that
are likely due to background processes. The exponent $\beta$ introduced
in the Valencia algorithm gives a handle to enhance or diminish the increase
of the $\frac{d_{ij}}{d_iB}$ ratio in the forward region, as shown in 
Figure~\ref{fig:distance_vlc}. Thus, we have a handle to {\em tune} 
the background rejection that is independent of the parameter $R$ that
governs the jet radius.

\section{Monte Carlo simulation}
\label{sec:simulation}

The performance of the different algorithms is compared
for $ t \bar{t}$ and $ZZ$ production at a linear $e^+e^⁻$
collider with $\sqrt{s} = $ 500~\gev. Samples are 
generated with WHIZARD~\cite{Kilian:2007gr}.
The response of the ILD detector~\cite{Behnke:2013lya}
is simulated with GEANT4~\cite{Agostinelli:2002hh}. 

\begin{figure*}[htbp!] 
{\centering 
 \includegraphics[width=0.45\textwidth]{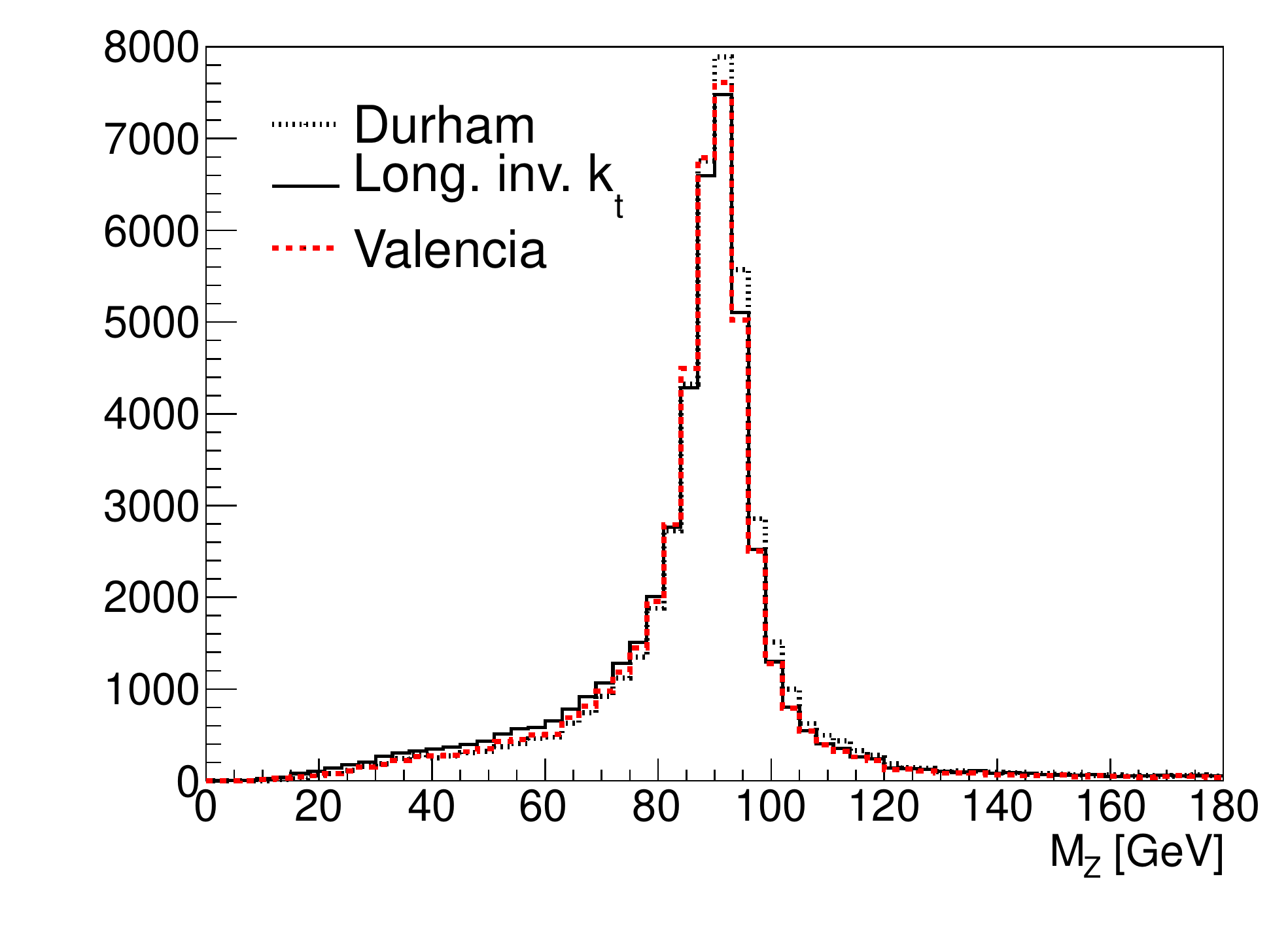}
  \includegraphics[width=0.45\textwidth]{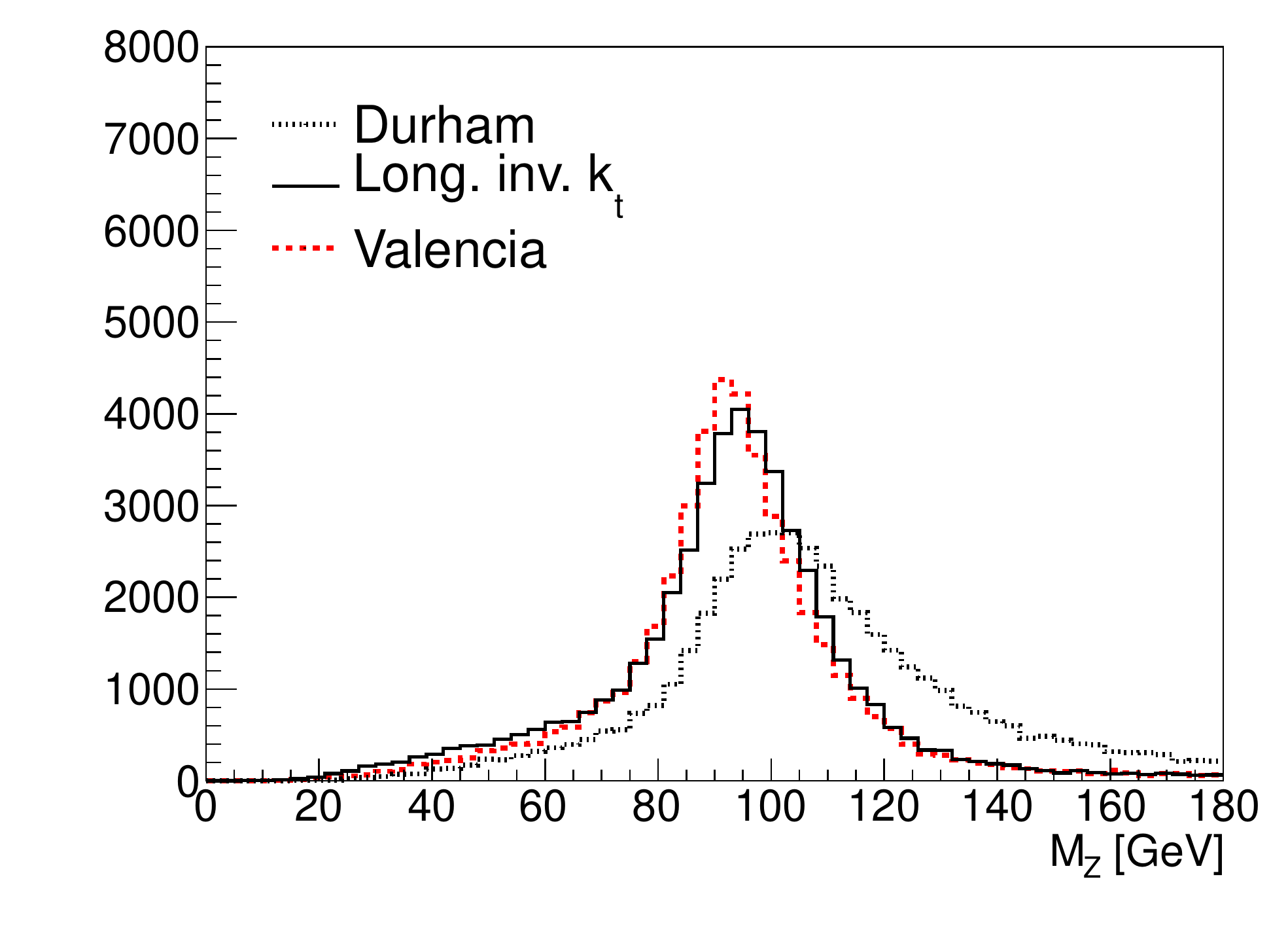}
  \caption{The reconstructed $Z$-boson mass distribution for $ZZ \rightarrow q\bar{q}q'\bar{q}'$ events at a 500~\gev{} CLIC. No backgrounds are added in the leftmost plot. The results on the rightmost plot correspond to the same events with the $\gamma \gamma \rightarrow$ {\em hadrons} background corresponding to 300 bunch crossings overlaid on the signal, where each bunch crossing contains approximately 0.3  $\gamma \gamma \rightarrow $ {\it hadrons} events.}
  \label{fig:clic_Z}
}
\end{figure*}

The background considered in this study is due to multi-peripheral 
$\gamma \gamma \rightarrow ${\it hadrons} 
production\footnote{A further source of background,
pair production from beamstrahlung photons is ignored 
in this discussion. }. The background events are overlaid on the signal
using a mechanism similar to that used for pile-up at the LHC. 
For a 500~\gev{} $e^+e^-$ collider less than one $\gamma \gamma \rightarrow $ 
{\it hadrons} events is produced per bunch crossing. 

The impact of the background on the output of the detector 
is quite different at CLIC and the ILC. 
At CLIC bunches are spaced by 500 picoseconds and detector systems are 
expected to integrate the background of a number of 
subsequent bunch crossings. In this study the background corresponding to 
a large number of bunch crossings is overlaid (300 for 500~\gev{} operation, 
60 for 3~\tev{}). The much larger bunch spacing 
at the ILC allows the detector to distinguish single bunch crossings, 
such that less than one  $\gamma \gamma \rightarrow $ {\it hadrons} event
is overlaid (on average) on each signal event.

In the event reconstruction the information of the tracking system 
and the calorimeters is combined to form particle-flow objects with the
Pandora~\cite{Marshall:2012hh} algorithm. 
In the CLIC studies particle flow objects are  
selected using a set of timing cuts, corresponding to the nominal selection
of Ref.~\cite{Marshall:2012ry}.

\section{Top quark pair production at a 500~\gev{} ILC}
\label{sec:resultsilc500}

We study the performance of several jet algorithms in the study of
$t\bar{t}$ production at the ILC of Ref.~\cite{Amjad:2013tlv}.
The Monte Carlo sample includes all six-fermion 
processes that produce a ``lepton + jets'' final state: $e^+e^- \rightarrow b \bar{b} l^\pm \nu_l q \bar{q'}$.

Reconstruction of the event involves charged lepton reconstruction and
removal of the corresponding energy, the reconstruction of exactly four
jets (exclusive jet clustering with $N = $ 4) and flavour tagging, 
described in detail in Ref.~\cite{Amjad:2013tlv}:
The two jets with poorest score in the b-tagging algorithm are combined 
to form the $W$-boson candidate. The hadronic top candidate is constructed 
by adding the remaining (b-)jet that minimizes a $\chi^2$ based
on the hadronic top quark candidate mass and energy, the b-jet energy 
in the top quark rest frame and the angle between $W$-boson and b-quark.

\begin{table}[h!]
        \begin{center}
	\begin{tabular}{l c c c c c}
		\hline
   RMS$_{90}$ [GeV]   & \hskip 0.5mm $ E_{4j}$ \hskip 0.5mm &  \hskip 0.5mm  $ E_{W} $ \hskip 0.5mm & \hskip 0.5mm $m_{W}$  \hskip 0.5mm &  \hskip 0.5mm $E_{t}$ \hskip 0.5mm  & \hskip 0.5mm $ m_{t} $  \hskip 0.5mm  \\  \hline
Durham           &   23.2   &   19.6   &   20.3   &   19.5 & 21.4 \\
$e^+e^-$ $k_t$    &   25.6   &   20.8   &   21.6   &   20.5 & 22.8 \\         
long. inv. $k_t$ &   21.7   &   18.4   &   18.9   &   18.4 & 20.1 \\
Valencia         &   21.4   &   18.0   &   18.8   &   18.2 & 20.0 \\ \hline                
	\end{tabular}
	\end{center}
	\caption{The Root Mean Square of the central 90\% of the events (RMS90)  for five observables reconstructed in $t\bar{t}$ events at a 500~\gev{} ILC: the energy of the system formed by the four jets, the energy and mass of the hadronic $W$-boson and the energy and mass of the hadronic top quark. }
	\label{tab:fsresultsresolution}
\end{table}

We consider four jet reconstruction algorithms: the Durham algorithm, the 
generic $e^+e^-$ $k_t$ algorithm with beam jets with $R=$ 1, the longitudinally
invariant $k_t$ algorithm with $R=$ 1.5 and the Valencia algorithm with $R=$1.2
and $\beta=$ 0.8. The choice of parameters correponds to the
optimal setting determined in a scan over a broad range of parameters.
The resolution of the measurements of the energy of the four jets, of 
the energy and mass of the hadronic $W$-boson and hadronic top quark 
candidate are given in Table~\ref{tab:fsresultsresolution}. 


The results show a clear advantage of the algorithms with a $d_{ij}/d_{iB}$ 
ratio that increases in the forward and backward region of the experiment. 
Even with the rather modest background level at the ILC the longitudinally 
invariant $k_t$ algorithm and the algorithm proposed in this Letter achieve 
a 10-15\% better resolution and a smaller bias than the $e^+e^-$ algorithms.

\section{Di-boson production at CLIC}
\label{sec:resultsclic500}

The $e^+ e^- \rightarrow ZZ$ process is studied in the CLIC environment 
to enable comparison with the first detailed studies of the impact of 
background on jet reconstruction at future lepton colliders in 
Ref.~\cite{Marshall:2012ry} and the CLIC CDR~\cite{Linssen:2012hp}. 

We select $e^+ e^- \rightarrow ZZ \rightarrow q\bar{q} q'\bar{q}'$ 
events. Events with $Z$-bosons emitted in the very forward direction 
(with polar angle $|\cos{\theta}| >$ 0.99), where the beam pipe may
have a profound impact are discarded, as well as events where the 
$Z$-bosons are very far from their mass shell ($|m(q\bar{q}) - m_Z| > $ 
30~\gev{}.

Exactly four jets are reconstructed and the di-jet combinations are
selected that minimize the following $\chi^2$:
\begin{displaymath}
\chi^2 = \frac{(E_{Z1} - E_{Z2})^2}{(250\gev{})^2} + \frac{(m_{Z1} - m_{Z2})^2}{(91 \gev{})^2} + \frac{\angle{(Z_1,Z_2)}}{(\pi)^2}.
\end{displaymath}

The $Z$ boson candidate mass distribution is shown in Figure~\ref{fig:clic_Z}. 
Numerical results are given in Table~\ref{tab:resultsclic}.
\begin{table}[h!]
        \begin{center}
	\begin{tabular}{l c c c }
		\hline
\multicolumn{4}{c}{$\sqrt{s} = $ 500~\gev{}, no background overlay} \\ \hline
   [ GeV ]            &  $m_Z$  & $\sigma_Z$ & RMS$_{90}$   \\ 
Durham           &  90.6   &   5.4   &   13.8   \\
long. inv. $k_t$ &   90.4   &   5.3   &   14.3   \\
Valencia         &  90.3   &   5.2   &   12.5   \\ \hline    
\multicolumn{4}{c}{$\sqrt{s} = $ 500~\gev{}, 0.3 $\gamma \gamma \rightarrow $ {\it hadrons} events/BX} \\ \hline
    [ GeV ]       &  $m_Z$  & $\sigma_Z$ & RMS$_{90}$    \\ 
Durham           &   101.1   &  13.6  &   28.8   \\
long. inv. $k_t$ &   95.1  &    10.9   &   17.9  \\
Valencia         &   93.1   &   10.2   &   17.1   \\ \hline 
	\end{tabular}
	\end{center}
	\caption{The center and width - from a Gaussian fit - of the reconstructed $Z$-boson mass peak in $ZZ$ events at a 500~\gev{} CLIC. The third column lists the RMS90 estimate. }
	\label{tab:resultsclic}
\end{table}









In the background-free case all three algorithms achieve a narrow $Z$-boson
mass peak. The impact of the overlaid background is rather pronounced for the
Durham algorithm. The peak position shifts by approximately 10~\gev{} and
broadens considerably. Both the longitudinally invariant $k_t$ algorithm
and the Valencia algorithm show considerably better performance under
these conditions.

\section{Conclusions}
\label{sec:conclusions}

We propose a jet algorithm that offers robust performance 
in the presence of the mild background levels expected at lepton colliders, 
while retaining the natural inter-particle distance criterion in the
[energy, angle] basis (as opposed to the [transverse momentum, rapidity] basis
of hadron collider algorithms).
The algorithm is further generalised with a variable exponent that allows
to tune the background rejection for the specific 
requirements of a given analysis.
We have benchmarked the performance of several algorithms in a full Monte Carlo
simulation studies of $t\bar{t}$ and $ZZ$ production at the ILC and CLIC. 
We find that the Valencia algorithm performs better than the sequential
recombination algorithms used at previous lepton colliders.
	
\section*{Acknowledgement}
The authors would like to thank Gavin Salam for helpful suggestions and guidance creating the plugin code and Bryan Webber for his careful reading of the manuscript.

\bibliographystyle{JHEP}
\bibliography{vlc_jets}	
	
\end{document}